\documentclass[14pt,a4paper]{article}
\usepackage{graphicx,amsmath,amsfonts}

\def \ba{\begin{eqnarray}}\def\ea{\end{eqnarray}}
\def\bc{\begin{center}}\def\ec{\end{center}}
\def\nn{\nonumber\\}

\title{\huge \bf Electromagnetic effects and
scattering lengths extraction from experimental data on $K\to 3\pi$
decays}

\author{\bf S.R.Gevorkyan\footnote{On leave of absence from
Yerevan Physics Institute}, D.T. Madigozhin, A.V.Tarasov,
O.O.Voskresenskaya\footnote{On leave of absence from Siberian
Physical Technical Institute} }

\begin{document}

\date{}

\maketitle
\bc Joint Institute for Nuclear Research, 141980 Dubna, Russia \ec

\begin{abstract}
The final state interactions in $K^\pm\to\pi^\pm\pi^0\pi^0$ decays
are considered using the  methods of non-relativistic quantum
mechanics. We show how to take into account the largest
electromagnetic effect in the analysis of experimental data, using
the amplitudes calculated earlier. We propose the relevant
expressions for amplitude corrections valid both above and below
the two charged pion production threshold $M_{\pi^0\pi^0}=2m_{\pi^\pm}$,
including the average effect for the threshold bin. These
formulae can be used in the procedure of pion scattering
lengths measurement from $M_{\pi^0\pi^0}$ spectrum.
\end{abstract}

The experiments DIRAC and NA48/2 at CERN SPS ~\cite{A,B} are
able to measure the $\pi\pi$ scattering lengths
difference $a_0-a_2$. Its value is predicted by Chiral
Perturbation Theory (ChPT)  with a high accuracy~\cite{CGL}
($a_0-a_2=0.265\pm 0.004$ in units of inverse pion mass.)
Thus the extraction of $a_0,a_2$ from experimental data with
comparable precision become an important task. Recently the NA48/2
experiment has discovered an anomaly (cusp) in the invariant mass
$M_{\pi^0\pi^0}$ spectrum of $\pi^0$ pairs from $K^\pm\to \pi^\pm\pi^0\pi^0$
decays. This anomaly is positioned at the charged pions production
threshold $M_c=2m_{\pi^\pm}$, and has been explained by N.Cabibbo~\cite{C}
as a result of charge exchange reaction $\pi^+\pi^-\to\pi^0\pi^0$ in
final state. It turns out that investigation of these decays below
the charged pions threshold $M_c$ allows one to extract the scattering
lengths with a precision not accessible in other current experiments.

Further development of the model has been done in the
works~\cite{CI,Gass,GPS}, where the pions final state interactions
were considered with the second order precision in scattering
lengths terms. Approach developed in ~\cite{CI} provides the possibility
to extract $a_0,a_2$ from experimental data~\cite{B} with a precision
compatible to theoretical prediction one. Nevertheless there are two
problems, requiring a special consideration, that can be crucial for
scattering lengths extraction from experimental data in the
vicinity of the threshold.

The first problem is the accounting for a higher orders in scattering
lengths terms and their impact on the value of $a_0,a_2$. The estimation
made in \cite{CI} leads to the corresponding $~5\%$ error for $a_0-a_2$.
In comparison with the precision of ChPT prediction and
with the experimental statistical error this is a
rather noticeable uncertainty, and its decreasing is very
desirable \footnote{More pessimistic view on impact of higher order
terms has given in ~\cite{GPS}}.

Another effect that has an impact on $a_0,a_2$ extraction from
experimental data is the electromagnetic interaction of pions
in the final state.

This issue is not a trivial task~\cite{I,G}, as Coulomb interaction
between charged pions\footnote{The photons radiation leads only to
smearing of Coulomb contribution and will be treated elsewhere}
below threshold $M_c$ leads to formation of bound $\pi^+\pi^-$
states (pionic atoms $A_{2\pi^{\pm}}$), and construction of them in
the framework of perturbation expansion is a doubtful issue.

To solve this problem we use the technique  of non-relativistic
quantum mechanics\footnote{The similar  approach to $K\to 3\pi$
decays was developed by V.Gribov~\cite{G1,G2}}. Leaving the strict
derivation for elsewhere, we cite here the main result for amplitude
of the decay under consideration
 \ba T&=&(1+iv_0f_0)T_0+2ivf_xT_+,\quad f_0=a_{00}/D;
 f_x=a_x/D,\nn
D&=&(1-iv_0a_{00})(1-2iva_{+-})+2v_0va_x^2 \label{f1} \ea

Here $m_0, m$ are the $\pi^0, \pi^{\pm}$ masses,
and $v_0=\frac{\sqrt{M^2-4m_0^2}}{2m_0};
v=\frac{\sqrt{M^2-4m^2}}{2m}$ are the neutral and charged pions
velocities respectively. $T_+$ and $T_0$ are the ``unperturbed''
matrix elements of $K^{\pm} \to \pi^\pm \pi^+ \pi^-$ and $K^{\pm}
\to \pi^\pm \pi^0 \pi^0$ decays. The inelastic $a_x$ and elastic
$a_{00},a_{+-}$  pion-pion scattering amplitudes in the isotopic
symmetry limit are ~\cite{CI,Gass} $a_x=(a_0-a_2)/3;
a_{00}=(a_0+2a_2)/3; a_{+-}=(2a_0+a_2)/6$. The replacement $a_i\to
f_i$ has small numerical impact on results of previous calculations
done according ~\cite{CI,Gass}, but is crucial for inclusion of the
electromagnetic interactions under threshold, where the formation of
bound states ($\pi^+\pi^-$ atoms)  take place.

The expression (\ref{f1}) includes all successive elastic and
inelastic interactions in the $\pi\pi$ system (to all orders in
scattering lengths). Their interaction with a spectator pion can be
taken into account at two-loop level~\cite{CI,Gass}.

As it was discussed earlier~\cite{GTV}, to include  the Coulomb
interactions in the frameworks of considered approach, it is enough
to make the simple replacement in (\ref{f1}): \ba v\to \tau
=iv-\alpha \left[\log(-2ivmr_0)+2\gamma+\psi(1-i\xi)\right];~~
\xi=\frac{\alpha}{2v}\label{f2}\ea where $\gamma=0.5772$,
$\alpha=1/137$ are Euler and fine structure constants, whereas
$\psi(\xi)=\frac{d\log{\Gamma(\xi)}}{d\xi}$ is digamma
function~\cite{AS}. The  parameter $r_0$ by its meaning is the
strong interaction radius, usually  taken as $r_0 \sim 1/m$. Later
on we cite it in all expressions, but as we checked, its impact on
the results of fitting is small.Strictly speaking the expression
(\ref{f2}) is valid in the region $vmr_0\leq 1$,however for
considered kinematics it is a rather well approximation.

To go under threshold one needs to perform the common replacement
$v\to i\tilde{v}$: \ba \tau=-\tilde{v}- \alpha
\left[\log(2\tilde{v}mr_0)+2\gamma+\pi\cot(\pi\tilde\xi)+\psi(\tilde\xi)\right];
\tilde{v}=\frac{\sqrt{4m^2-M^2}}{2m} \label{f3} \ea Due to
asymptotic behavior  $\psi(z) \sim log(z)$ this expression is finite
at threshold $M_c$.Moreover it describes both all bound states
(pionium atoms) and electromagnetic interactions leading to unbound
states under threshold~\cite{GTV}.

Above the threshold the expression (\ref{f2}) acquires the imaginary
part. Using the well known ~\cite{AS} relations  for digamma
function we obtain
 \ba \tau&=&Re\,\tau+i Im\,\tau \nn
Re\,\tau&=&-\alpha \left[\log(2mr_0)+\gamma+
\xi^2\sum_{n=1}^{\infty}\frac{1}{n(n^2+\xi^2)}\right],\nn
Im\,\tau&=&\frac{\pi\alpha}{1-e^{-2\pi\xi}}\label{f4} \ea

The above expressions allows one to calculate  the electromagnetic
corrections in wide kinematic region. For definiteness let us consider
the fitting procedure in  NA48/2.
In order to extract the values of scattering lengths $a_0,a_2$
~\cite{B}, the function \ba F(M^2) = N \int f(a_0,a_2,m^2)
\Phi(M^2,m^2) dm^2 \label{f5} \ea was fitted to the experimental
mass spectrum. Here $f$ is the theoretical distribution
from~\cite{CI}, whereas $\Phi$ describes the experimental setup
resolution and acceptance and obtained by Monte-Carlo simulation.
Integral is replaced with a sum, and decay probability $f$ is
calculated in the middle of every $M_{\pi^0\pi^0}$ bin with a width
$\delta M^2 = 0.00015 (GeV/c)^2$. The bins are centered in such a
way, that the threshold $4m_{\pi\pi}^2$ is placed exactly in the
middle of one of them (here we will call this bin as the ``central''
one).

This procedure is precise enough as long as one considers only strong
interactions in final state. But the electromagnetic interactions lead
to the sharp peak at the threshold, thus for the central bin the averaging
of theoretical predictions must be done more carefully.

So it is convenient to consider the behavior of the electromagnetic
corrections in three separate regions. Up to the lower bound of
central bin due to smallness of $\xi$ in this region from (\ref{f3})
we get
\ba
\tau&=&\frac{\pi\alpha}{2}cot(\frac{\pi\alpha}{2\tilde{v}})+
\alpha\left(log(2\tilde{v}mr_0)+
\gamma-\frac{\alpha^2}{4\tilde{v}^2}\varsigma(3)\right);\nn
\varsigma(3)&=&\sum_{1}^{\infty}{\frac{1}{n^3}}=1.201\label{f6}
\ea

This expression for $\tau$ have to be used there instead of the
velocity formula for charged pions $\tilde{v}=\frac{\sqrt{|M^2-4m^2|}}{2m}$.

Above the charged pions production threshold, beginning from the top
bound of central bin the interference between direct $T_0$ and
charge exchange $T_+$ terms appears (unlike in
~\cite{C}). As a result to take into account the electromagnetic
effects in this region one has to add the interference term \ba -4\alpha
a_x\left(log(2mr_0)+\gamma+\frac{\alpha^2}{4\tilde{v}^2+\alpha^2}+
\frac{1}{2}log(\tilde{v}^2+(\frac{\alpha}{3})^2)\right)T_0T_+
\label{f7} \ea to the square of matrix element.

For these regions the averaging procedure (just a value in the bin's
center) exploited in ~\cite{B} is precise enough.

As to the central bin, where the electromagnetic corrections come
from two states of $\pi^+\pi^-$  pairs (bound or unbound) the more
accurate averaging is in use. Confining  to the second order terms
in scattering lengths, as it is done in previous
consideration~\cite{CI}, substituting the relevant velocities
(\ref{f3},\ref{f4}), integrating the square of amplitude
(\ref{f1}) in central bin and dividing result by the bin width
$\delta M^2$, we obtain the mean value of correction to the square of
matrix element in the central bin in the form \ba
\bar{T^2}&=&\frac{\pi\alpha^3\varsigma(3)}{2v_0 v_t^2}T_+^2-4\alpha
a_x\left(log(2v_tmr_0)+\gamma-0.5\right)T_+ T_0;\nn
v_t&=&v(4m^2+\frac{\Delta M^2}{2});~~v_0=v_0(4m^2) \label{f8}\ea

The first term  describes all bound states (pionium atoms).
As to the second term it is a result of interference between the
direct "unperturbed" production amplitude $T_0$ and electromagnetic
interactions leading to unbound states.Let us note that in
(\ref{f8}) we cite only the main electromagnetic corrections in
central bin. We estimated the possible contributions from unbound
states ($ \sim T_+^2$ ) and interference between direct amplitude
$T_0$ and bound states ($\sim T_+T_0$) in central bin.They are
 small (less then 1\% in comparison with main terms in
(\ref{f8})) and so can be safely neglected.

Numerically the contribution of the first term (pionium
atoms) from (\ref{f8}) into the central bin is near $~2.5\%$  of
``unperturbed'' decay probability, that coincides with the
prediction evaluated from ~\cite{S}. Simple estimates show that the
interference of direct amplitude $T_0$ with unbound Coulomb part
(second term in (\ref{f8})) is significant and gives approximately
the contribution of same order as bound states one.

Now it becomes clear why the fit done in ~\cite{B} with the pionium
contribution as a free parameter leads to the Coulomb contribution,
that is about two times larger, than can be provided by solely bound
states.

In conclusion let us stressed that at present we have the recipe
allowing to introduce the Coulomb effects into  the expressions for
the square of decay amplitude (taken, for example, from \cite{CI}),
averaged correctly over every bin, that can be multiplied by phase
space factor, convoluted with the experimental acceptance matrix
$\Phi$ and compared with the experimental data.\\
 We are grateful to V.~Kekelidze and J.~Manjavidze who draw our
attention to the problem and advice during all the work.

\end{document}